\documentclass{article}
\usepackage{amsmath}
\usepackage{amsfonts}
\usepackage{amssymb}
\usepackage{mathtools,harmony,upgreek}
\usepackage{color,accents,stmaryrd}
\usepackage{mathrsfs}
\usepackage{amssymb}
\usepackage{amsmath}
\usepackage{dsfont}
\usepackage{indentfirst}
\usepackage{slashed}
\newcommand{\cl}{C \kern -0.1em \ell}

\newcommand{\p}{\partial}

\newcommand{\m}{\mu}
\newcommand{\n}{\nu}

\newcommand{\s}{\sigma}

\newcommand{\g}{\gamma}

\newcommand{\be}{\begin{equation}}
\newcommand{\ee}{\end{equation}}
\newcommand{\ba}{\begin{array}}
\newcommand{\ea}{\end{array}}
\newcommand{\f}{\frac}

\newcommand{\beq}{\begin{eqnarray}}
\newcommand{\eeq}{\end{eqnarray}}
\newcommand{\ot}{\otimes}

\newcommand{\om}{\Omega}

\renewcommand{\b}{\beta}

\renewcommand{\a}{\alpha}

\renewcommand{\b}{\beta}
\renewcommand{\a}{\alpha}
\definecolor{clearblue}{rgb}{0,0.5,0.9}

\definecolor{clearblue}{rgb}{0,0.5,0.9}
\definecolor{orange}{rgb}{1,0.5,0}

\begin{document}

\markboth{K. P. S. de Brito}
{Review on Spinor Fields and Applications in Physics}

%
%
\begin{center}
{\bf REVIEW ON SPINOR FIELDS AND APPLICATIONS IN PHYSICS}
\vspace{2cm}

{\it K. P. S. DE BRITO
\vspace{.5cm}
\footnote{kelvyn.paterson@hotmail.com}}

{\small Department of Physics, Federal University of ABC,\\
Santo Andr\'e, SP
Brazil} 
%
%
\vspace{1cm}
\end{center}
\begin{abstract}

A review about spinor fields is presented, constructing a outlook through the last century. Spinor was explored in many contexts more and more in the last decades. Besides this, more papers about this issue has been produced in the last decade than in the others before it. As examples, classifications of spinor fields on the bulk and on compactified 7-manifolds done by the author are revised and some results about spinor fields on this contexts are explored, like generating of brane by singular spinors in the bulk, invariants on five-dimensional black holes and spectral decomposition of a quantum field on compactified dimensions at low energies. 
\end{abstract}

\tableofcontents








\section{Introduction}

Many studies around Clifford algebras, and his useful derived objects called spinor, has been motivated by physical and mathematical reasons. Besides this, this issue was explored in many contexts. Through the history many progress took place, since remote times. We know that since the beginning of civilization, the set of numbers and the set of mathematical objects has been increased at facing new problems and challenges in human society. 

The history maybe starts with the Hamilton's attempt in construct a geometric algebra with three dimensions and his discovery about  the quaternions as an extension of the complex number set. After this, other extensions was constructed. Among them, Grassman built the exterior algebra. Clifford enumerated his quadratic geometric algebra  in a table of classification, which have periodicity eight, as it was shown by Cartan. Cayley explored  a interacted process to weaken properties in extension of algebras, like real number are extended to complex number losing  the duality property ($\bar{z}=z$) and complex to quaternion losing the commutativity \cite{Baez}. Geometric constructions inside physical and mathematical scenarios was facilitated with those new algebraic tools, beside this new intuitions and insights about such themes arises, because his more geometric nature. Like example, spin groups was used to represent rotation by Lipschitz, Riesz wrote the exterior product using the Clifford product. 


It is interesting to remember that a Clifford algebra starts from a vector space and a quadratic form and that the ``spinor field'' idea, which is used to describe fermions and started to be explored in the begin of the last century \cite{Purk}, is constructed from such algebra as a double covering of the special orthogonal group $SO(n)$ \cite{lou2,Lawson}. Besides, they were classified on Minkowski space by Lounesto \cite{lou2} and so on many exploration was done at this branch. It is interesting to salient that one the firsts to mention the term ``Lounesto's classification'' was Romo in a paper of 1993 \cite{8}, as it is accused in the Google Scholar. Thenceforth, many aspects have been explored.

\section{Goal}
Here we are interested in a short review on the last studies about Clifford algebra and spinor fields, in their classification on some spaces, like the bulk and compactified spaces, such examples are important to fermions dynamics in quantum gravity. Besides this, a spinor quantum field is introduced to compactified spaces at low energies; a type of spinor field on the bulk degenerate in another that generate the brane; and the classification of spinor fields on the bulk is corroborated by a verification that the components of the bilinear form on the bulk fill out all the invariants construct with the spinor field.

\section{Bilinear Covariants  and Spinor classification on Minkowski space}

Bilinear covariants are constructed with spinor fields and can be used to classify them, through some constraints named Fierz identities and can be associated to observable of quantum fields. This classification was done in Minkowski space by Lounesto, where he listed his six classes of spinors, being three named regular spinors and the others three named singular spinors, specifically: flagpole, flag-dipole and dipole \cite{lou2}. A similar work was done in five-dimensional Lorentzian spaces, in particular in the bulk \cite{BR}, and in seven-dimensional Riemannian\cite{BBR} and Lorentzian \cite{BR2016} spaces, as important examples can be considered the seven-sphere ($S^7$) and the seven-dimensional Anti-de Sitter space ($AdS_7$) respectively. Those classification, other few results about spinor fields and their applications on physics are reviewed here.

Given a spinor field $\psi$ on the Minkowski space, bilinear covariants are written as follows:
\begin{subequations}
\begin{eqnarray}
\sigma &=& \bar{\psi}\psi,\label{sigma}\\
\mathbf{J}&=&{{\rm J}_{\a }\theta ^{\a }=\bar{\psi}\gamma _{\a }\psi\, \theta
^{\a}},\label{J}\\
\mathbf{S}&=&S_{\a \b }\theta ^{\a}\wedge\theta^{ \b }=\tfrac{1}{2}i\bar{\psi}\gamma _{\a
\b }\psi \,\theta ^{\a }\wedge \theta ^{\b },\label{S}\\
\mathbf{K}&=& K_{\a }\theta ^{\a }=i\bar{\psi}\gamma_{5}\gamma _{\a }\psi
\,\theta ^{\a},\label{K}\\\omega&=&-\bar{\psi}\gamma_{5}\psi\,,  \label{fierz.}
\end{eqnarray}\end{subequations}
where $\gamma_5:=i\gamma_0\gamma_1\gamma_2\gamma_3$ is the volume element  in Minkowski spacetime
\cite{livro}. They can be used to describe the observables of the electron \cite{dirac,Dirac}: the current density corresponds to $\mathbf{J}$, $\mathbf{S}$ is the spin density, the chiral current is denoted by $\mathbf{K}$ and $\omega$ gives spinor indications under the CPT symmetry. Obviously, all them are bilinear on the spinor field $\psi$ and some constraints result from them, through the Fierz identities, which are written with such bilinear covariants \cite{lou2}:
\begin{equation}\label{fifi}
-(\omega+\sigma\gamma_{5})\mathbf{S}=\mathbf{J}\wedge\mathbf{K},\qquad\mathbf{K}\lrcorner \mathbf{K}+\mathbf{J}\lrcorner\,\mathbf{J}
=0=\mathbf{J}\llcorner\mathbf{K},\qquad
\mathbf{J}\,\lrcorner\, \mathbf{J}=\omega^{2}+\sigma^{2}\,.  
\end{equation}
These constraint, that are valid only for regular spinors, can be generalised to encompass the singular spinor case also, transforming the constraints above in the following expressions with a much better appearance \cite{cra}: 
\beq\nonumber
Z\gamma_{\mu}Z=4J_{\mu}Z,\qquad Z^{2} =4\sigma Z,\qquad iZ\gamma_{\mu\nu
}Z=4S_{\mu\nu}Z,\nonumber\\ 
-Z\gamma_{5}Z=4\omega Z,\qquad  iZ\gamma_{5}\gamma_{\mu}Z=4K_{\mu}Z.
\label{boom}
\eeq
From these constraints, and other not represented here, comes the classification of the spinors obtained by Lounesto. Many aspects around the Fierz identities in the relativistic field was explored in some papers recently: All them are reduced to only one equation through extended Cartan map \cite{10}. Inversely, Fierz identities are used to reconstruct the spinors \cite{14}. Completeness and orthogonality relations are obtained through Fierz identities \cite{6,7}, where it is constructed an equivalence between tensor and spinor representations. Computations of constraints involving the exotic term of Dirac equation \cite{exotic,daSilva:2016htz,daRocha:2016bil}, a general comprehension about the Elko spinor together its relation with the Dirac spinor \cite{jmp07}, a action in supergravity for each type of spinor \cite{elko}, some aspects on the Dirac operator \cite{Vassilevich:2015soa} 
  and a result about Hawking radiation obtained from black string tunneling \cite{bht,Cavalcanti:2015nna}.

Summarizing, the Fierz identities provide constraints to the bilinear covariants, and consequently to the spinor fields $\psi$, splitting all the set of spinor fields on the Minkowski space into such six classes described below, besides other three classes with null current density, which can represent ghost states \cite{EPJC}:  
\begin{eqnarray}
{\bf regular}\;\left\{
\begin{array}{cc}
1)\;\omega\neq0,\;\;\;  \sigma\neq0,\\
\label{Elko11}
2) \;\omega = 0,\;\;\;
\sigma\neq0,\label{dirac1}\\
3)\;\omega \neq0, \;\;\;\sigma= 0,\label{dirac21}
\end{array}
\right\} & - \;\;\boxed{\;\;{\bf J}\neq 0,\;\;\mathbf{K}\neq 0, \;\;\;\mathbf{S}\neq0  \;\;}& \\
 {\bf singular}\;\boxed{\omega=\sigma=0,\;\;{\bf J}\neq 0},&\;
\left\{\begin{array}{cc}
4) \;\;\mathbf{K}\neq 0, \;\;\;\mathbf{S}\neq0  \;\;\text{(flag-dipole spinors)}\quad\qquad\\
\!\!\!\!\!\! 5) \;\;\mathbf{K}=0,\;\;\; \mathbf{S}\neq0 \;\; \text{(flagpole spinors)}\quad\qquad\\
\!\!\!\!\!\!\!\! 6)\;\; \mathbf{K} \neq 0,\;\;\; \mathbf{S}=0
\;\; \text{(dipole spinors)}\quad\qquad
\end{array}\right. & 
\\
{\bf ``ghosts''}\;\boxed{\omega=\sigma=0,\;\;{\bf J}= 0},&
\hspace{-4cm}\left\{\begin{array}{cc}
7) \;\;\mathbf{K}\neq 0, \;\;\;\mathbf{S}\neq 0 \\
\!\!\!\!\!\! \; 8) \;\;\mathbf{K}=0,\;\;\; \mathbf{S}\neq 0 \\
\!\!\! 9)\;\;\mathbf{K} \neq 0,\;\;\; \mathbf{S}=0
\end{array}\right. & 
\end{eqnarray}

All classes with ${\bf J}\neq 0$ was found to have at least a possible physical representative. This result was fruit of exploration of those spinor fields in many contexts by many authors. Flagpoles was considered in the context of cosmology recently \cite{shank} and proposed to be dark matter candidate in many scenarios \cite{exotic,daSilva:2016htz,daRocha:2016bil,alex,lee2,lee1}.
%
 Majorana and Elko \cite{DVA,lee1,Jardim:2014xla,Liu:2011nb,shank,lee2,bht,daRocha:2005ti,elko,hopf,jmp07,Cavalcanti:2014uta,HoffdaSilva:2009is,Fabbri:2011mi,Elkodirac,alex,horv1,wund,Pereira:2014pqa,Basak:2014qea,Alves:2014qua,Fabbri:2014foa,S.:2014dja,Agarwal:2014oaa,Pereira:2014wta,m1} spinors fields are examples of flagpoles realisations. In the same sense, Weyl spinors are examples of dipole spinors and, recently, it was found a missing example of flag-dipole in ESK  models of gravity like solutions of the Dirac equation
in a background with $f(R)$-torsion \cite{esk}.  

\section{Bilinear covariants of Spinor Fields on Arbitrary  Dimensions and Signatures}

The bilinear covariants written with the spinor fields in Minkowski space can be extended to spaces $M$ of arbitrary dimensions $n$ and signatures $p+q$. Observables of the spinor $\psi$ are, in the same way that in Minkowski space, associated to their bilinear covariants, as shown in the previous section. In such construction, the dual spinor is identified to the conjugation $\bar{\psi}=\psi^\dagger\gamma^0$. This is generalised, if it is considered a hermitian operator $A=A^\dagger$, in this way the dual spinor is written as $\bar{\psi}=\psi^\dagger A^{-1}$ and the bilinear covariants can be expressed by the following expressions \cite{bt,Pesk,BBR}:

\begin{eqnarray*}
&B_0=\bar{\psi}\psi,\\
&\vdots\\
&B_{\a_1\ldots\a_k}=\f{1}{n!}\bar{\psi}\gamma_{\a_1}\cdots\gamma_{\a_k}\psi;\;k<n,\\
&\vdots\\
&B_{1\ldots n}=\f{1}{n!}\bar{\psi}\gamma_1\ldots\gamma_n\psi .
\end{eqnarray*}

As it was previously mentioned, the classification of the spinor fields in arbitrary dimensions and signatures comes from the Fierz identities built with their respective bilinear covariants. However, besides this, it is necessary to introduce the homomorphisms $D$ and $J$, which provides structures for the space of spinors. The homomorphism $D$ is named real structure and implements the complex conjugation, for other side $J$ is the complex structure that is equivalent to multiplication by the imaginary unit
\beq
\psi^*:=D(\psi),\qquad J^2=-{\bf 1}.
\eeq
 Depending on the dimension and signature of the space, the structures $D$ and $J$ can constraint the bilinear covariants \cite{1,moro,Babalic:2013fm,bonora,face,lazaroiu}.

\section{Spinor Fields on seven-dimensional Riemmanian spaces}

It is intended a classification of spinor fields on Riemannian spaces of seven dimensions, in particular on $S^7$, that is an important case in compactified space of 11-dimensional supergravity ($AdS_4\times S^7$) \cite{SUGRA,SUSYBr,engl}. The real and complex structures provide constraints on spinor fields, that are fixed by the Fierz identities. In such way, Majorana spinor on such spaces, that are the only class of spinor existent in the literature, are extended through complexification, resulting in other two new non-trivial classes of spinor fields \cite{1,BBR}.

The importance of this classification resides in the physical aspects of the results that will come from the exploration of spinor fields on seven dimensional compactified spaces \cite{1,lou2,ced,SUGRA,SUSYBr,BBR}, in the same way that spinor fields classification in Minkowski space in accord to Lounesto has empowered unexpected discover and a large path of new physical approaches and new particles \cite{alex,lee2,lee1,daRocha:2005ti,exotic,esk,Cavalcanti:2014wia,daSilva:2012wp,daRocha:2007sd,m1}.  

Given a Majorana spinor $\psi$ and a bilinear form $B$. The number of non-null bilinear covariants can be reduced through the constraint coming from the property of being Majorana spinor and from the symmetry of $B$. The first gives that $B(\psi,\g^A\psi)$ can be different from zero, only if $|A|$ is even, where $A$ is a multi-index and $|A|$ is its degree. The second one gives $B(\psi,\gamma^A\psi)=(-1)^{|A|(|A|+1)/2}B(\psi,\gamma^A\psi)$ can be non-null, only if $|A|=0,3,4,7$. Hence, $B(\psi,\gamma^A\psi)$ can be zero, except if $|A|=0,4.$ Then, it remains only the possible non-vanishing bilinear covariants $\phi_0=\bar{\psi}\psi$ and $\phi_4=\bar{\psi}\gamma_{ijkl}\psi$. If the Fierz identities are imposed, they are reduced to the following expression:
\beq
(\phi_4+1)\circ(\phi_4+1)=8(\phi_4+1),
\eeq
where $\circ$ is the Clifford product. This fix the inequalities for the bilinear covariants and it constraints the Majorana spinor to only one class, as it is knew in the literature \cite{1,BBR}. 

The complex ($J$) structures, which is equal to the volume element $\g_1\cdots\g_7$ in seven dimensions, and the real ($D$) structure are used to construct other three bilinear forms:
\beq
B(\psi,J\gamma_A\psi),\quad B(\psi,D\gamma_A\psi),\quad B(\psi,JD\gamma_A\psi).
\eeq
It can be verified that the constraints found previously are identical to the constraints correspondent to second bilinear form above. It is possible to infer also that the first and third form are equivalent. Because this, without redundancies, it remains only the constraints coming from the first form. In this way, a selection rule and the symmetry of the bilinear form provide that the bilinear covariant $B(\psi,J\gamma_A\psi)$ is different from zero if, respectively, $|A|$ is odd and $B(\psi,J\gamma_A\psi)=(-1)^{|A|(|A|+1)/2}B(\psi,J\gamma_A\psi)$, i.e., $|A|=0,3,4,7$. Summarising, $B(\psi,J\gamma_A\psi)$ vanishes, except if $|A|=3,7.$ The case $|A|=3$, that refers to the bilinear form $(\psi,J\gamma_{ijk}\psi)e^{ijk}$ that has a correspondent in the literature, because this form can be associated to the three-form torsion gauge, which is being to much extensively explored in the last decade \cite{SUGRA,Gimb,eng4,akiv,logi09,engl,SUSYBr,ced}.
 However, there is a Hodge duality $(\star\;\cdot\;\;)$ that relates the two important bilinear forms: $\f{1}{|A|!}B(\psi,J\gamma_A\psi)e^A=\star\f{1}{|A'|!}B(\psi,\gamma_A'\psi)e^{A'} $, where $|A|+|A'|=7.$ Because this fact, the constraints coming from the bilinear covariant $B(\psi,J\gamma_A\psi)$ do not provide any new information to the classification of the spinor field $\psi$ \cite{1,BBR}.

Finally, a general classification of the spinor fields on seven-dimensional Riemannian spaces is obtained, if the bilinear covariant $B(\psi,\gamma_A\psi)$ is complexified into a more general bilinear form, as follows:
\beq
\beta(\psi,\gamma_A\psi)=B(\psi_R,\gamma_A\psi_R)-B(\psi_I,\gamma_A\psi_I)+i\left[B(\psi_R,\gamma_A\psi_I)+B(\psi_I,\gamma_A\psi_R)\right],
\eeq
where $\psi_R$ denotes the real part and $\psi_I$ denotes the imaginary part. In this case, some cancellations enable null and non-null bilinear covariant values. Indeed, those guarantee the following more general classification for arbitrary spinor fields on seven-dimensional Riemannian spaces:
\begin{subequations}
\beq & \upvarphi_0=0,\quad\upvarphi_4=0,\label{c11}\\
& \upvarphi_0=0,\quad\upvarphi_4\neq0,\label{c12}\\
& \upvarphi_0\neq0,\quad\upvarphi_4=0,\label{c13}\\
& \upvarphi_0\neq0,\quad\upvarphi_4\neq0\,.\label{c14}\eeq
\end{subequations}

It is noticed two new classes of spinor fields: one regular and other singular. These new classes can result in new possible candidates to fermion particles at low energy in spontaneous compactifications of $D=11$ supegravity \cite{SUGRA} and, besides this, it arises interesting possibilities about physical solutions in the massless sector of gauged $N=8$ supergravity theory \cite{engl,BBR}. Fot other side, added to the contexts of the Killing spinors and parallelizable seven-sphere \cite{ced,eu1}, the dynamic of those spinor fields in each class enumerated above can be explored beginning with the Lagrangian terms, which in a quaternionic realization assume the form:
\beq
M&=&tr(\bar{\psi}\psi)\\
M_\alpha&=&tr(\bar{\psi}\g_\mu\psi)\\
K_{\beta\gamma}&=&tr(\bar{\psi}\g^\alpha\g_{\beta\gamma}\p_\alpha\psi)\\
K_\delta &=&tr(\bar{\psi}\g^\m\g_\delta\p_\m\psi)
.\eeq
Because our classification, we can infer that only the massive term contributes. Hence, this implies that in last instances in low energies does not exist dynamics to fermionic fields on the seven-dimensional compactified spaces \cite{BBR, Top}. As it was shown by Toppan \cite{Top}, the terms of the Lagrangian that need to be considered depend on the realisation taken in consideration. It is interesting to observe that the quaternionic realisation above is the most plausible, because the other realisations: real and octonionic, prohibit at first the existence of kinetic terms. In such way that in such realisations only appears non-Weyl mass terms of the type $M_\perp = tr(\bar{\psi}_+\psi_-+\bar{\psi}_-\psi_+)$ \cite{Top}.


\subsection{Some considerations about Compactifications} 

The fermionic fields classification that we proposed play a fundamental role at compactified models in $D=11$ supergravity. The first compactification models was proposed in the begin of last century by Klein, who improved a model of unification in five-dimensional spaces \cite{duff86,nast,Duff}, where was proposed one the first tentatives in unify the gravity force with other force. The improvement of Klein consisted into decomposing the scalar field in a Fourier serie through compactifying a dimension into a circle \cite{duff86}. Similar procedures can be done in high-dimensional spaces. For example, in a two-dimensional compactification with sphere form, the scalar field $\varphi$ can be decomposed in a spherical harmonic serie, and go on for high-dimensional spheres. This gives the following interpretation: ``For $q$ compactified dimensions, a non-massive theory in $4+q$ dimensions results in a massive theory in $4$ dimensions'' \cite{duff86}, as represented by the equation below.
\beq
\square_{4+q}\varphi=(\square+\nabla_q)\varphi=(\square+m^2)\varphi.
\eeq
The next step would be to quantize the field, for this intuit, look that the fermionic quantum field in Minkowski space takes free value of momentum (modes) \cite{Pesk}:
\beq
\psi(x)=\int\frac{d^4 p}{(2\pi)^4}\sum_i\left(a^s_{\bf p}u^s(p)e^{-ip\cdot x}+b^{s\dagger}_{\bf p}v^s(p)e^{ip\cdot x}\right).
\eeq
In this way, consider the state $\Psi$ on the compactified space $M^4\times K_q$, where $M^4$ is the space-time and $K_q$ is a q-dimensional compactified space. It is decomposed like $\Psi(x,y)=\psi(x)\Psi_q(y)$. Hence, observe that in $\Psi_q$ there is creation and annihilation operators associated to each harmonic spherical (mode) in the discrete spectra. The field factor $\Psi_q$ is denoted at low energy on $K_q$ by:
\beq 
\Psi_q(y)=\sum_{I_n}\left[Y_n^{I_n}(y)u_{I_n}a_{I_n}+\left(Y^{I_n}_n(y)\right)u^*_{I_n}b_{I^*_n}\right],
\eeq 
where $I^*_{l,m}=I_{l,-m}$ is a multi-index and $a_{I_n},b_{I^*_n}$ are operator valued coefficients.
\section{Spinor Fields on the Bulk}

Given their importance on studies in brane theories, it is intended here to review the classification of spinor fields on the bulk \cite{brane}. A similar procedure to shown above in four and seven dimensional spaces is done in four plus one, specifically, in the bulk \cite{BR}. Consider a bilinear form $B$ on the bulk, the deal with spinor fields on the bulk is more elaborated than in seven dimensions, because it is necessary to introduce quaternionic structures $J_i$, that satisfy the relations \cite{1,BR}
\beq
J_i J_j=-\delta_{ij}+\epsilon_{ij}^kJ_k.
\eeq
Nevertheless, the proceture to classify spinor fields on the bulk is much similar to approached in the previous section at seven dimensions. The constraints come from the form symmetries and selection rule, and classes are fixed in a classification by the Fierz identities. Then, from the symmetry of $B$, the bilinear covariants are constrained by
$$
\phi_A:=B(\psi,\g_A\psi)e^A=(-1)^{\frac{|A|(|A|-1)}{2}}B(\psi,\g_A\psi)e^A.
$$
This implies that $B(\psi,\g_A\psi)$ is zero, except if $|A|=0,1,4,5$.

The quaternionic structures are used to construct other three bilinear forms $B(\psi,J_i\g_A\psi)$, from whose symmetries comes other constraints for the spinor fields on the bulk. Explicitly, the bilinear form is constrained by $$
\hat{\phi}_A^i:=B(\psi,J_i\g_A\psi)e^A=(-1)^{\frac{|A|(|A|-1)}{2}+1}B(\psi,J_i\g_A\psi)e^A
,$$ 
that vanishes, except if $|A|=2,3$. From where arises other possibly non-vanishing six bilinear covariants. However,  the number of bilinear covariants is cut by half, because the Hodge dualities $\phi_A=\star \phi_{A'}$ and $\hat{\phi}_A^i=-\star \hat{\phi}_{A'}^i$, where $|A|+|A'|=5$. Then, only the following bilinear covariants remain: $\phi_0,\;\phi_1$ and $\hat{\phi}_2^i$, where $i=1,2,3$, which are the only ones possible to be non-null. Besides this, the Fierz identities
\beq
\phi_1\circ\phi_1-\sum_i\hat{\phi}_2^i\circ\hat{\phi}_2^i=7(\phi_0)^2+6\phi_0\phi_1,\\
\phi_1\circ\hat{\phi}_2^i+\hat{\phi}_2^i\circ\phi_0+\sum_{jk}\epsilon_{ijk}\hat{\phi}_2^j\circ\hat{\phi}_2^k=6\phi_0\hat{\phi}_2^i
\eeq
 fix them to form only one class, as found in the literature \cite{1,BR}. This class can be extended through a kind of complexification to encompass arbitrary spinor fields on the bulk
\beq
\beta_k(\psi,\g_A\psi)=B(\psi,\g_AJ_1\psi)-iB(\psi,\g_AJ_3\psi).
\eeq 
Through this process, it is found other six non-trivial class of spinor fields on five-dimensional Lorentzian spaces like the bulk, where three are regular spinor fields and the other three are singular \cite{1,BBR,BR}, as it is shown below:
\begin{subequations}
\beq
{\bf regular}\;-\; \phi_0\neq 0,\left\{\begin{tabular}{cc}
$\quad \phi_1\neq 0,$& $\left\{\begin{tabular}{c}
$\quad\hat{\phi}^i_2\neq 0,$\\ $ \quad\hat{\phi}^i_2=0,$\label{c11}
\end{tabular}\right.$\\
$\quad \phi_1= 0,$& $\left\{\begin{tabular}{c}
$\quad\hat{\phi}^i_2\neq 0,$\\ $ \quad\hat{\phi}^i_2=0,$\label{c11}
\end{tabular}\right.$
\end{tabular}\right.\\
{\bf singular}\;-\;\phi_0= 0,\left\{\begin{tabular}{cc}
$\quad \phi_1\neq 0,$& $\left\{\begin{tabular}{c}
$\quad\hat{\phi}^i_2\neq 0,$\\ $ \quad\hat{\phi}^i_2=0,$\label{c11}
\end{tabular}\right.$\\
$\quad \phi_1= 0,$& $\left\{\begin{tabular}{c}
$\quad\hat{\phi}^i_2\neq 0,$\\ $ \quad\hat{\phi}^i_2=0,$\label{c11}
\end{tabular}\right.$
\end{tabular}\right.
\eeq
.\end{subequations} 
It is important to salient that the contributing terms of the Lagrangian for each class of spinor fields depends on its realization \cite{Top, BR}, but this is beyond the scope of this paper.

\subsection{Degeneracy
 of the flag-dipole Spinor into a dipole spinor generating of brane}

Spinor constrained to the brane contained in the bulk must obey the classification for spinor fields in Minkowski space \cite{lou2,brane,BR}. In this way, it can be analyzed if a spinor field in a given class generate brane or does not, calculated the coupling constants and his correspondent energy density, and how this occur \cite{brane,BR}. Given a scalar field $\varphi$ massless on a Randall-Sundrum background
\beq
ds^2=exp(-2A(r))\eta_{\mu\nu}dx^\mu dx^\nu +dr^2,
\eeq
the graviton modes can be calculated and their masses \cite{brane}. If a fermionic spinor field represented by $\psi=(a(r),0,b(r),0)^\intercal$ is added to this system, the spinor components can be solved through the Dirac and Einstein equations
\beq
{R_a}^A-\frac{1}{2}{e_a}^A R=\chi {T_a}^A+{e_a}^A\Lambda,\qquad [i\slashed{D}-m+\lambda(\bar{\psi}\psi)]=0,
\eeq
and be expressed in terms of the extra dimension and of the coupling constant. Additionally, the energy density can be computed, resulting in $T_0^0=-2$ \cite{brane,BR}. 

If instead to consider the spinor field given above, we do different and we consider a two-components flag-dipole expressed by $\psi=(a(r),0,0,b(r))$ and the metric $ds^2=\varphi^2(r)(\eta_{\m\n}dx^\m dx^\n)-dr^2$, we surprisingly found a null energy-momentum tensor. Because, such flag-dipole does not generate brane and does not have existence. However, from the constraint ${T_{\bar{0}}}^1=0=\varphi'(r)a(r)b(r)$, we can infer that the flag-dipole degenerate into a one-component dipole spinor field $\psi=(a(r),0,0,0)^\intercal$, whose component and the correspondent metric can be computed \cite{Gimb,brane,BR}
\beq
\varphi(r)=\varphi_0 exp(\pm r\sqrt{\Lambda/6}),\qquad a(r)=a_0 exp(\mp r\sqrt{2\Lambda/3}).
\eeq 

\subsection{Invariant on axisymmetric black holes}

Consider a five-dimensional axisymmetric and stationary black hole and a observable Killing vector $\xi^\m$ on it. The results obtained in the begin of this section can be corroborate here through this example. There is a current vector $J^\m$ associated and equal to this Killing vector. If a null Killing vector in the event horizon is translated from the black hole neighborhood to the spatial infinity it became a time Killing vector, but the one-form current density is preserved in the {\it f\"unfbein} basis $e^A={e^A}_\m dx^\m$ \cite{Mei1,Mei2,MeiBH,BR}. If it is imposed the contraint to the spinor field $\psi$: $(\g^0\pm \g^5)\psi=0$, it is expressed in components like $\psi=(\a_1,\a_2,\a_1,\a_2)^\intercal$. In a given representation to the gamma matrices on the bulk with signature four plus one
\beq
\g^0=i\s^1\ot {\bf 1}_2,\quad \g^k=-\s^2\ot\s^k,\quad \g^4=\s^3\ot{\bf 1}_2,
\eeq
the Killing vector is represented by:
\beq
\xi^\m=\p_t+\om_1\p_{\phi_1}+\om_2\p_{\phi_2},
\eeq
where $\phi_1,\phi_2$ and $\om_1,\om_2$ denote the related two azimuth angles and two azimuth angular velocities from the angular coordinates of the sphere $S^3$ correspondent to the five-dimensional black hole. 

Remember that there are three bilinear forms on five-dimensional Lorentzian spaces that can be non-vanishing and can represent observables associated to the spinor field $\psi$: 
\beq
\phi_0&=&\bar{\psi}\psi,\qquad\qquad\qquad\;-\;{\bf normalization}\\
\phi_1&=&\bar{\psi}\g_\m\psi e^\m,\qquad\qquad\;-\;{\bf current \;vector}\\
\hat{\phi}_2^i&=&\f{1}{2}\left(\bar{\psi}J_i\g_{\m\n}\psi\right)e^{\m\n}\;\;\,-\;{\bf dual\; spin}.
\eeq 
The constraint imposed on $\psi$ permits to write all the bilinear form in function of the components of $\psi$. This implies in redundancies on the bilinear forms components, which the different ones reduces to the following four terms:
\beq
|\a_1|^2\pm|\a_2|^2,\qquad \a_1^\dagger\a_2\pm\a_2^\dagger\a_1
.\eeq
It is important to say that these components fill out all the invariants that can be constructed on the five-dimensional Lorentzian spaces with the spinor field $\psi$ constrained \cite{BR}.
 \section{Conclusion}
 
 In this work, we reviewed some aspects of the studies about spinor fields on some spaces like on Minkowski space \cite{lou2}, on seven-dimensional Riemannian spaces \cite{BBR}, for example $S^7$, and on five-dimensional Lorentzian spaces \cite{BR}, for example the bulk. Besides a general expression for bilinear covariants that are listed on spaces of arbitrary dimension and signatures \cite{BBR}. Summarizing, the author obtained in previous papers two new non-trivial classes of spinor fields \cite{BBR}, in particular, on seven-dimensional compactified spaces, which play a very crucial rule on $D=11$ supergravity, beside six new non-rivial classes of spinor fields on the bulk \cite{BR}.
 
 Studies about compactification permit us to write the quantum field to fermionic particle at low energies on compactified spaces. Besides this, a view around brane enabled us to verify that the flag-dipole spinor degenerate into the dipole spinor, which generate brane in the bulk. For other side, an analysis on the cosmological point view through axisymmetric black hole permitted us infer, given the constraint for the spinor field: $(\g_0\pm \g_5)\psi=0$, that bilinear covariants components fill out all the invariants, which can be built with a spinor field on five-dimensional Lorentzian space.

\section*{Acknowledgments}

The author thanks to the very prolific discussions with Rold\~ao da Rocha and to the UFABC, CAPES and CNPq (grant No. 150882/2017-3) grants  for support.

\footnotesize

\end{document}